# Computational role of sleep in memory reorganization

Kensuke Yoshida[1,2], Taro Toyoizumi[1,2]

[1] Laboratory for Neural Computation and Adaptation, RIKEN Center for Brain Science, 2-1 Hirosawa, Wako, Saitama 351-0198, Japan,

[2] Department of Mathematical Informatics, Graduate School of Information Science and Technology, The University of Tokyo, 7-3-1 Hongo, Bunkyo-ku, Tokyo 113-8656, Japan,

Correspondence should be addressed to Kensuke Yoshida at kensuke.yoshida@a.riken.jp and Taro Toyoizumi at taro.toyoizumi@riken.jp

## Abstract

Sleep is considered to play an essential role in memory reorganization. Despite its importance, classical theoretical models did not focus on some sleep characteristics. Here, we review recent theoretical approaches investigating their roles in learning and discuss the possibility that non-rapid eye movement (NREM) sleep selectively consolidates memory, and rapid eye movement (REM) sleep reorganizes the representations of memories. We first review the possibility that slow waves during NREM sleep contribute to memory selection by using sequential firing patterns and the existence of up and down states. Second, we discuss the role of dreaming during REM sleep in developing neuronal representations. We finally discuss how to develop these points further, emphasizing the connections to experimental neuroscience and machine learning.

# Highlights

- Slow waves might selectively consolidate memory by imposing low-dimensional neuronal activities
- Up and down states in slow waves could differentially contribute to memory consolidation
- Dreaming might form efficient cortical representations of memories

# Introduction

Sleep is an essential physiological state conserved across a variety of species, including nematodes, flies, and mammals. Sleep is considered to reorganize memory of awake experiences into an efficient form [1–3], inducing consolidation and assimilation of experiences and inspiring inference from learned relationships [4–7]. Given its evolutionary conservation, sleep is expected to play an essential role in learning that cannot be achieved during wakefulness.

Some sleep characteristics have been well explored theoretically. During sleep, our brain lacks sensory inputs but internally generates neural activity. Mimicking these characteristics, in the Helmholtz machine composed of recognition and generative connections between the input and hidden layers, learning of recognition connections is driven during the sleep phase by neural activity caused by its generative model instead of real sensory inputs, while learning of generative connections is driven during the wake phase by sensory inputs [8,9]. This suggests that the combination of wakefulness and sleep might contribute to extracting hidden input structures. Another sleep characteristic is that firing patterns about awake experiences are replayed during sleep [10,11]. Mimicking this, more recent studies suggested that replays, which are not necessarily created by generative models but are sampled from the memory buffer (storage of previous inputs), are effective for preventing catastrophic interference and stabilizing learning [12,13]. While these studies have demonstrated the utility of memory replays in artificial machine learning methods, other studies begin to propose the computational roles of more detailed sleep characteristics.

Sleep is divided into non-rapid eye movement (NREM) and rapid eye movement (REM) sleep with different characteristics. NREM sleep is characterized by slow waves, low-frequency (0.5 - 4.0 Hz) components of EEG and local field potential, which accompany synchronous transitions of cortical neurons between up states with higher membrane potential and down states with lower membrane potential (Fig. 1A) [14]. Slow waves have been considered to have causal influences on memory reorganization, especially when they are temporally coupled with neuronal reactivation of awake experiences [4,10,15–19]. While the role of the homeostatic synaptic regulation during sleep in memory consolidation has been investigated [20–22], the function of slow waves has not been sufficiently addressed theoretically. On the other hand, REM sleep is dominated by high-frequency neuronal activities as with wakefulness but accompanies dreaming and hallucinatory experiences created in our brain [23]. The involvement of REM sleep in associative thinking and creativity has been suggested mainly in human studies [3,24]. In addition, a recent study in rodents suggested that the communication from the medial entorhinal cortex to the anterior cingulate cortex during REM sleep is critical for the emergence of inference [7]. Therefore, how REM sleep and dreaming promote learning needs to be theoretically investigated.

In this review, we discuss the possibility that NREM sleep selectively consolidates memory and, based on it, REM sleep creates efficient neuronal representations. First, we consider that two features of slow waves: sequential firing patterns along the propagation of slow waves and the existence of up and down states may contribute to the selection for memory consolidation. Second, we discuss the role of dreaming during REM sleep in developing efficient representations of memories. Dreaming created by our brain might help to explore and compare more global features of memory patterns. Thus, the NREM and REM sleep characteristics may synergistically enhance learning.

# Slow waves selectively consolidate memory within low-dimensional neuronal space

The transitions from down to up states of slow waves in cortical neurons propagate as a traveling wave, which evokes sequential firing patterns in neurons (Fig. 1A). This might be involved in memory selection by consolidating memories that are compatible with low-dimensional waves [25,26]. In the setting of [26], sequential stimulations of neurons in the cortex during wakefulness form a one-dimensional chain of synaptic weights that facilitates sequential activity in one direction (Fig. 1B). During subsequent sleep, slow waves promote replays of the learned sequence and further strengthen these synapses (Fig. 1B). Awake-like asynchronous firing patterns are not as efficient as synchronous slow waves in consolidating the learned sequence. Coherent waves expedite learning by efficiently promoting stereotypical neural activity patterns. These results suggest that memory can be strengthened using slow waves. A subsequent study further suggested the effects of slow waves on protecting against catastrophic forgetting in a similar setup [27]. When the model learns one direction of neural activity along layers and its reverse direction using the typical asymmetric spike-timing-dependent plasticity (STDP) time window, their effects interfere with each other. However, spontaneous replays toward both directions induced by slow waves allow sequences with opposing directions to coexist by assigning distinct subsets of neurons for the two directions (Fig. 1C). Another study also pointed out that replays induced by slow waves could improve visual classification [28]. These studies proposed that slow waves efficiently select memories to be consolidated by promoting spontaneous replays compatible with their low-dimensional waves. This could also help to realize replays without remembering all inputs in a memory buffer, which is important in machine learning [12].

Another model proposed the computational benefits of slow waves on goal-directed behavior [29]. This study suggested that the combination of slow waves and a reward-modulated update rule of synaptic weights enables neuronal networks to efficiently learn polysynaptic paths that involve multiple synapses in between (Fig. 1D). With awake-like asynchronous neuronal activity patterns, intact sequences to a distant target neuron rarely appear. However, with the aid of slow waves that promote sequential activity spreading to distant neurons, the reward-modulated synaptic update rule could induce the strengthening of synaptic weights along the paths. Further, repetitive activation of neurons by slow waves enables an efficient search over various candidate paths and could replace an initial detour path with a shortcut path that increases reward by achieving shorter behavioral latency (Fig. 1D). This result implies that slow waves can consolidate memory in a form that is useful for task solving.

In summary, these models show that slow waves can facilitate selective memory consolidation by confining replays within low-dimensional neuronal space. The restriction to the low-dimensional space might introduce inductive bias for assuring spatial continuity of neural activity and expediting learning. Although these methods have not yet been implemented in practical machine learning tasks, they constitute interesting computational hypotheses about the benefit of sleep in learning.

## Multiple states in slow waves coordinately reorganize memory

A recent theoretical study proposes that synaptic plasticity depends on states during slow waves [30]. This study adopted a normative approach based on the idea that biological systems have evolved to achieve optimality in some aspects [31,32] and especially investigated the possibility that synaptic changes are optimized for maximizing information transmission between neurons. This idea is theoretically formulated by the information maximization (infomax) synaptic plasticity that changes synaptic strength for increasing the mutual information (quantifying "amount of transmitted information" in information theory) between presynaptic and postsynaptic spike trains (Fig. 2A) [33]. The infomax synaptic plasticity differs in up and down states of slow waves and also depends on their spatial wavelength. First, the infomax synaptic plasticity is biased toward synaptic weakening when the baseline postsynaptic firing rate is high in up states (Fig. 2B). This effect is because the cost of strengthening a synapse is greater than its benefit for information transmission in the up states where frequent background inputs from many synapses degrade information transmission by each synapse (i.e., at a low signal-to-noise ratio). Second, the infomax synaptic plasticity predicts that global up and down states should induce more synaptic strengthening than local up and down states, respectively (Fig. 2B). This effect is again explained by the weakening bias of the infomax synaptic plasticity at higher baseline postsynaptic firing rate. In the excitatory-inhibitory network model that exhibits slow waves, surrounding neural activity increases and decreases the baseline firing rate of neurons at the center in their down states and up states, respectively (Fig. 2C). The cortical interaction is excitation dominated in down states but inhibition dominated in up states because inhibitory neurons are assumed to have a steeper rise of gain with input than excitatory neurons. Hence, the activity of the surrounding neurons suppresses that of central neurons in global up states but facilitates it in local down states (Fig. 2C). This surround suppression property is a reminiscent feature of stabilized supralinear networks [34–36] that has gained several experimental supports. This indicates that the global up and down states have lower firing rates than the local up and down states, respectively. Hence, the infomax synaptic plasticity predicts that synaptic changes should be positively biased in both up and down states if surrounding cortical areas take the same state as the center (i.e., in global up and down states). Indeed, these predictions are consistent with two recent rodent experiments [18,37].

Furthermore, integrating synaptic changes in down and up states might exert a delicate selection mechanism for memory consolidation. A recent theoretical study suggested that the formation of cell assemblies (i.e., coactive neuronal subpopulations with strong connections in between) depends on inhibition strength. It proposed that a network with dominant excitatory interactions rapidly forms nonspecific cell assemblies that recruit even non-stimulated neurons, while that with dominant inhibitory interactions slowly forms more specific ones that selectively recruit only stimulated neurons [38]. Considering the stabilized supralinear property of the slow-wave model described above [30], similar differences may be observed between the down and up states of NREM sleep–down states may promote the formation of nonspecific cell assembly rapidly compared with up states. While many experimental studies about memory reorganization have focused on neuronal activities during up states [18,19,39], a recent experimental study reported that neuronal activities during down states are also related to hippocampal ripple activities and, thus, memory consolidation [40]. Hence, up and down states of slow waves with distinct synaptic plasticity and excitatory-inhibitory balance may play complementary roles in memory consolidation. For example, up states might reorganize normal memory with higher specificity, while down states might promote the consolidation of crucial memory, such as life-threatening events, rapidly with lower specificity. It is an interesting future research topic to examine the more specific division of roles in these states.

## Learning representations by dreaming during REM sleep

Dreaming, often accompanied by REM sleep, is apparently not just a replay of past experiences but more a creative experience generated by our brain [41]. Although the detailed mechanism and content of dreaming are unknown, it is expected to be created by combining our previous experiences. How these imaginations could improve the brain's computation constitutes an attractive research topic. Previous studies have proposed that REM sleep might contribute to creatively discovering hidden associations [3,24]. In this section, we consider the hypothesis that dreams during REM sleep promote exploring more global structures of the input space by generating imaginary combinations of experiences to find efficient representations that reflect a possible relationship between remote memories (Fig. 3A). We review neuroscience studies about dreaming and state-of-the-art representation learning methods in the following paragraphs to explore potential ties.

One recent study [42] addressed the role of dreams by constructing a cortical implementation of generative adversarial networks (GANs), a successful generative model in machine learning by training a 'generator' to fool a 'discriminator' that learns to distinguish fake and real data [43]. (Fig. 3B). The model proposes a learning rule of feedforward (from lower to higher) and feedback (from higher to lower) synaptic connections. During wakefulness, both feedforward and feedback pathways learn similarly to the autoencoder [44] by compressing sensory input into high-level representation and reconstructing the original sensory input based on the high-level representation. During NREM sleep, low-level sensory representation is generated from replays by the feedback pathways with added noise, and the feedforward pathways learn to reconstruct the original high-level representation. During REM sleep, which is termed 'adversarial dreaming,' feedback pathways convert 'creative dreams,' i.e., fused memory, into low-level sensory representation, and feedforward pathways process it to judge if it is a dream. The feedforward pathways learn to discriminate correctly ('discriminator'), and the feedback pathways learn to fool them ('generator'). The NREM stage adds robustness to the high-level representation, and the REM stage improves the object-identity classification of sensory input. This study implies that the synthesized data explored by dreaming during REM sleep could contribute to forming representations.

Other studies in machine learning also provide hints about the possible roles of dreaming. It has been pointed out that negative samples in contrastive learning would be biologically implemented during sleep [45]. Contrastive learning is another powerful representation learning method in machine learning [46], based on the idea that similar data (termed 'positive sample') should be embedded close to each other, whereas dissimilar data (termed 'negative sample') should not. The augmented version of the original data and neighboring epochs in the video frames are often used as positive samples, and other irrelevant data are often used as negative samples. Although positive samples are considered easily realized in the brain by extracting neighboring frames from our experiences, negative samples seem relatively difficult to implement. Dreaming, in which inconsistent episodes are neighboring in time series, might provide combinations of data that work as negative samples. Therefore, dreaming might improve the representation by exploring the combinations that should be compared. Note that a recent paper suggested the possibility that contrastive learning could be realized during NREM sleep in the framework of [42] (see [47] for details).

In summary, dreaming might contribute to forming efficient neural information representations by exploring and comparing more diverse combinations of experiences than during NREM sleep. Such representation learning may have commonalities with machine learning methods [42,43,45], but the range of distortions involved might go beyond typical engineering setups today. Another potential difference with machine learning might lie in the alternation of NREM and REM sleep several times in one night. NREM sleep consolidates some recent memories with slow waves, and REM sleep may find hidden relationships between these memories and ostensibly unrelated others during dreaming. The prior memory selection by NREM sleep in iteration might help form relationships between memories, avoiding false relationships.

## Future perspective

Recent theoretical models suggest that slow waves during NREM sleep regulate memory selection, and dreaming during REM sleep form representations of selected memories. Theoretical models serve as an excellent guide to testing this hypothesis, bridging the gap between experimentally observed sleep characteristics and their computational roles. First, it would become possible to chronically track neuronal representation changes by using recently-developed methods for recording many neuronal activities at a high temporal resolution [48]. Such continual recording over sleep periods would elucidate whether underlying changes in neuronal circuits during NREM and REM sleep are consistent with our hypothesis with the aid of theoretical models. Second, the potential performance of the methods incorporating sleep characteristics needs to be studied. Although many sleep characteristics discussed in this review are only implemented for proof-of-principle demonstration, their computational benefits in more practical tasks should be also addressed in the future. In summary, cross-disciplinary investigation of sleep functions will be a key to understanding how the brain learns and discovering beneficial components for improving machine-learning methods in the future.

## Figure legend

Figure 1: Slow waves might selectively consolidate memory by arranging cortical replays
   A. Cortical neurons transit between up and down states synchronously with slow waves. During up and down states, the membrane potential of each neuron is higher and lower, respectively. Replays of spike sequences (orange) are promoted, especially during the transition from down to up states (gray shadow).
   B. Theoretical model in [26]. Sequential stimulations in a one-dimensional chain of cortical neurons (from 'A' to 'E') strengthen the synaptic weights in one direction (shown in red) during training, which is further strengthened during post-learning sleep.

C. When two opposing patterns (one from 'A' to 'E' and one from 'E' to 'A') are trained, replays during slow waves allow two sequences to coexist by assigning distinct subsets of neurons (shown in magenta and green) for the two directions [27].

D. Slow waves might improve computation by learning appropriate poly-synaptic paths. Such effects could be beneficial for finding the shortest path from the start neuron (S) to the target neuron (T) in a recurrent network (shown in green) [29].

Figure 2: Model of the state-dependent synaptic plasticity based on information maximization.

A. The model suggests that sleep might induce optimal synaptic changes for maximizing mutual information between presynaptic and postsynaptic spike trains. The infomax plasticity is formulated by increasing mutual information $I$ under the constraint of synaptic weight cost $\Phi$.

B. The model predicts that baseline firing rates modulate optimal synaptic plasticity. Therefore, the optimal synaptic plasticity and memory reorganization depend on the up and down states of global and local slow waves. These differences are consistent with previous experimental findings [18,37].

C. Distinct effects of the surrounding inputs during up and down states in an excitatory-inhibitory network model. Due to the supralinear transfer functions of neurons, inputs from surrounding populations suppress the excitatory firing rates in the center during up states (shown as an orange line), which is known as surround suppression, while they facilitate the excitatory firing rates in the center during down states (shown as a blue line). This property, the so-called stabilized supralinear property, has been experimentally supported [34–36]. As a result, excitatory firing rates are expected to be higher in local up and down states than global up and down states, respectively. Furthermore, this property implies the possibility that the externally driven neural activity spreads to surrounding populations during down states while it is restricted to the center during up states, which might lead to different spatial spreads of cell assemblies.

Panels A and B are modified from [30].

Figure 3: Dreaming might contribute to forming efficient cortical representations

A. Dreaming synthesized during REM sleep might play a role in forming efficient cortical representations. Dreaming might contribute to discovering possible associations and distinctions between memories.
B. The model suggests the possibility that forming cortical representations is affected by 'adversarial dreaming' during REM sleep and 'replay' during NREM sleep [42]. During REM sleep, inspired by GANs, the model trains a discriminator (feedforward projection) to distinguish whether it is a dream and a generator (feedback projection) to fool it. During NREM sleep, the feedforward projection is trained to reproduce representations in a higher cortical area under the condition that feedback transformations are perturbed. During wakefulness, the feedforward and feedback projections are trained to reproduce original inputs provided in a lower cortical area.

## Conflict of interest statement

The authors declare no conflicts of interest.

## Acknowledgments

This study was supported by RIKEN Center for Brain Science, Brain/MINDS from AMED under Grant No. JP15dm0207001 (T.T.), KAKENHI Grant-in-Aid JP18H05432 (T.T.) and JP21J10564 (K.Y.) from JSPS, RIKEN Junior Research Associate Program (K.Y.), and Masason Foundation (K.Y.).

## References and recommended reading

Papers of particular interest, published within the period of review, have been highlighted as:
*       of special interest
**      of outstanding interest

(annotations for selected literature)
[5] (*)
They investigated a time period when synaptic plasticity in the hippocampus and cortex is critical for memory consolidation by developing an optical tool for selectively erasing LTP within a specific time window. Memory consolidation requires the LTP during sleep, on the same day as learning in the hippocampus but on the second day in the anterior cingulate cortex.

[7] (*)
The study investigated the role of sleep in acquiring the relationship between memories using a transitive inference task in rodents. They showed that artificial enhancement of the communication from the medial entorhinal cortex to the anterior cingulate cortex during REM sleep promoted the inference after insufficient learning.

[26] (*)
The study proposed that spike sequence is replayed and consolidated along with up state propagation and its inducing synaptic plasticity by using a thalamocortical network model of slow waves including the STDP.

[27] (*)
They proposed that sleep phases, including slow waves, can prevent catastrophic forgetting. When a spatiotemporal firing sequence and its reverse firing sequence are learned continually in the model, the first sequence is forgotten without sleep but protected with sleep by allocating each sequence to different synapses.

[29] (**)
They showed that traveling waves and reward-modulated STDP can find useful polysynaptic paths. The model discovers a shortcut path in a recurrent network and organizes polysynaptic paths appropriate for solving tasks in a multi-layer feedforward network.

[30] (**)
This study proposed a normative model of sleep function with a close relationship to experimental findings. The study suggested that synapses change during NREM sleep to maximize neural information transmission. Optimal synaptic plasticity for the infomax is dependent on baseline firing rates, which is consistent with the experiments on synaptic plasticity and memory reorganization during up and down states of slow waves at distinct spatial scales.

[42] (**)

They proposed that dreaming might be beneficial for learning cortical representations. In the model, REM sleep conducts learning related to discrimination between 'creative dreams' and original memory, which is inspired by GANs, whereas NREM sleep conducts reconstruction of original memory from 'replays' with added noise. They suggested that the combination of REM and NREM sleep contributes to forming latent representations.

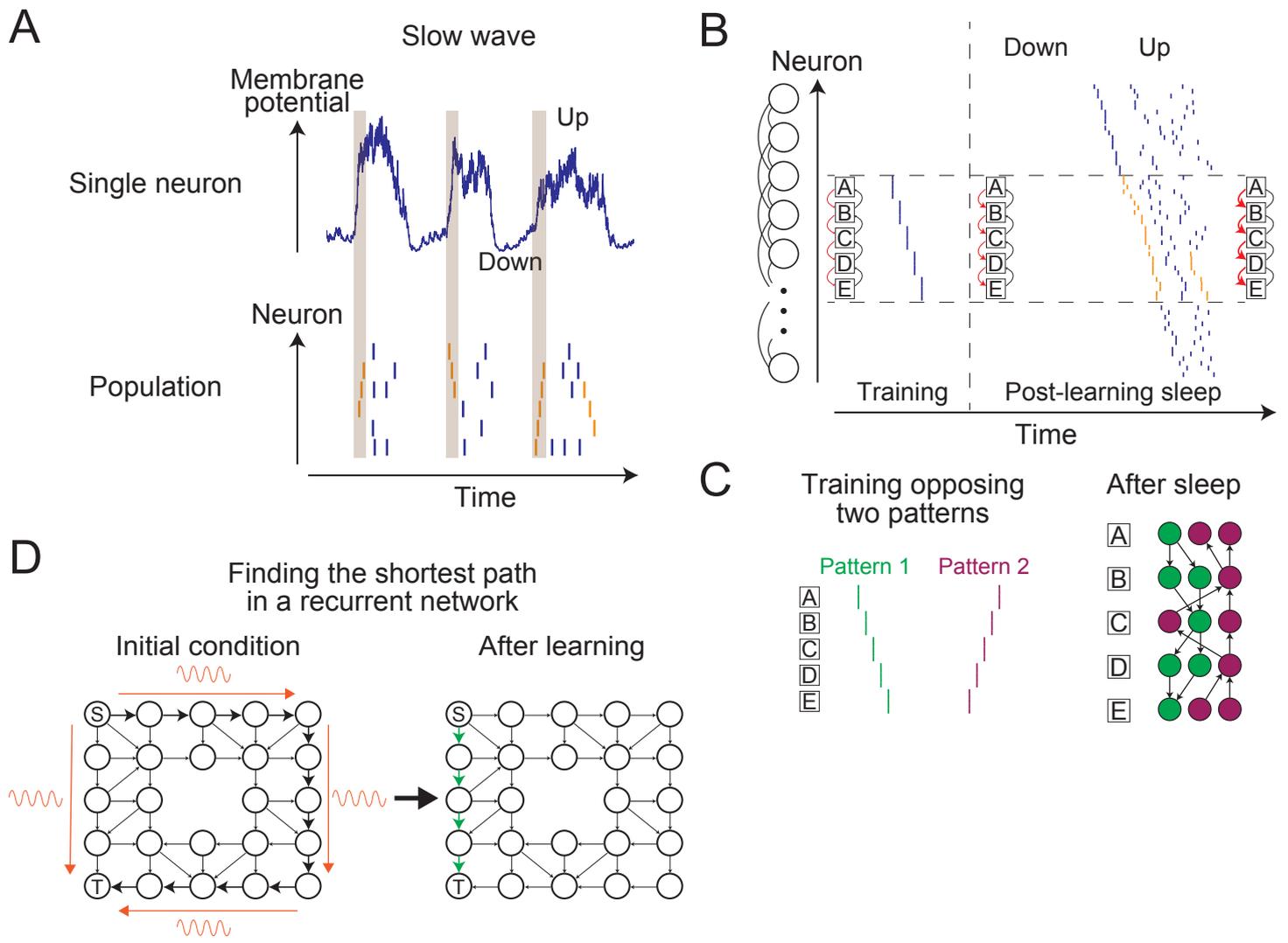

Figure 1

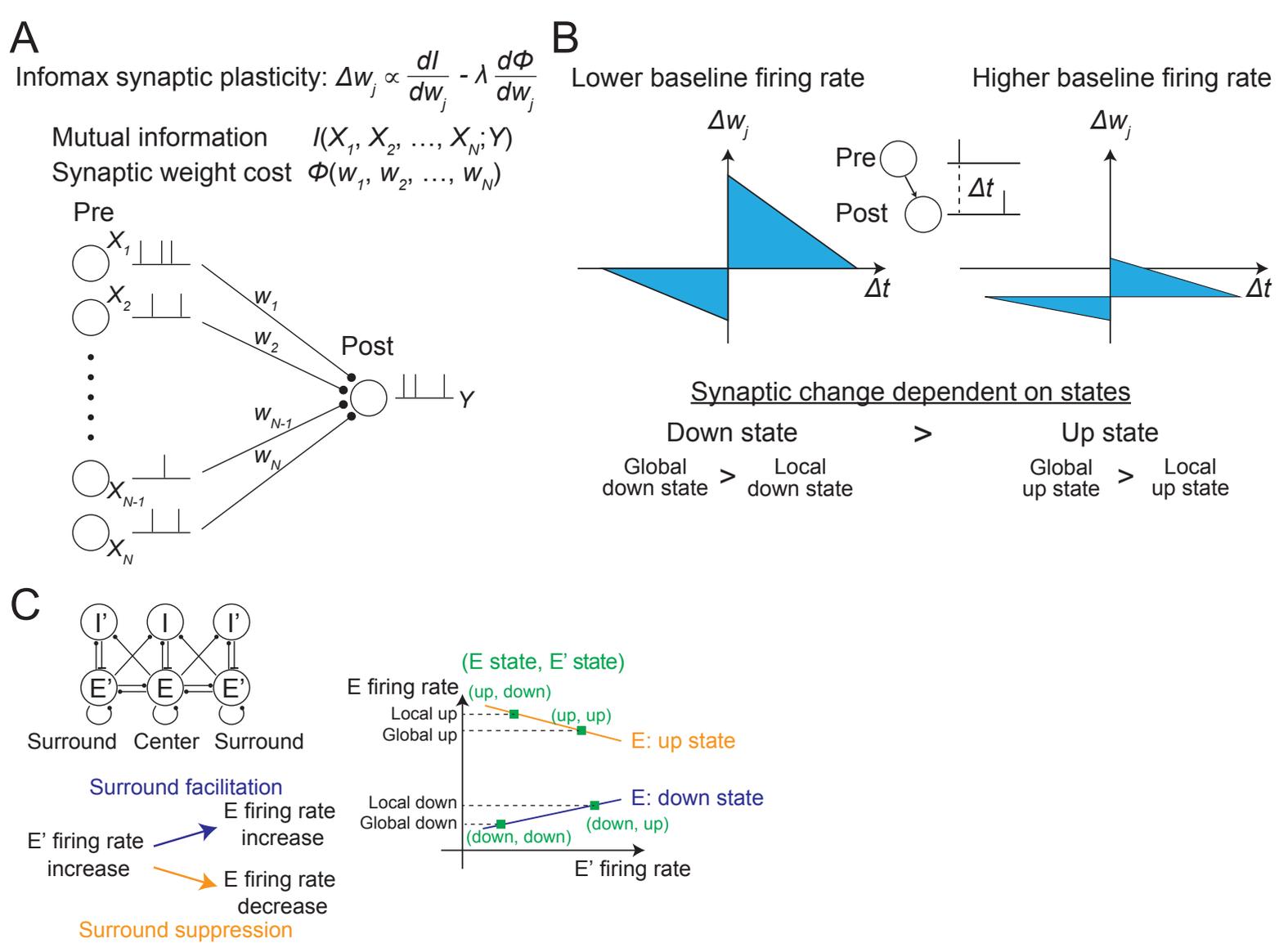

Figure 2

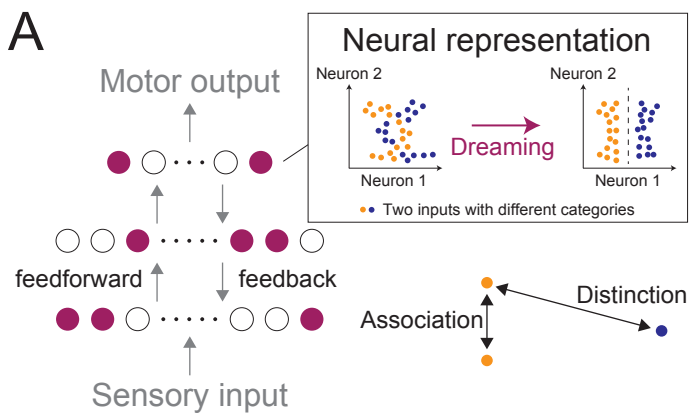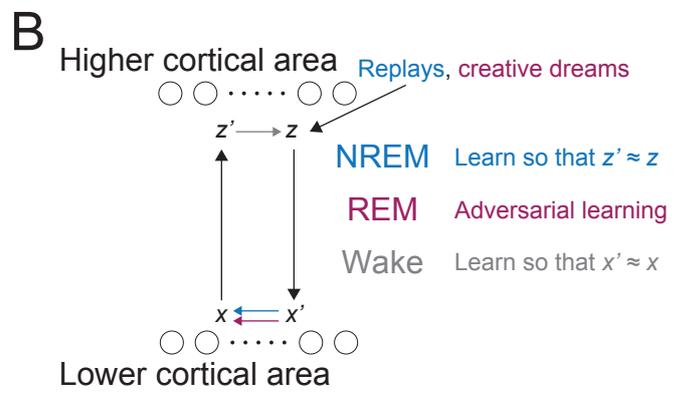

Figure 3